\documentclass[11pt]{article}
\newcommand{\Comment}[1]{{}}
\usepackage{amsfonts,amsmath, amsmath,amssymb,slashed}
\usepackage[textwidth = 430 pt, textheight = 630 pt]{geometry}

\Comment{\usepackage{color}
\definecolor{MyDarkBlue}{rgb}{0.15,0.15,0.45}
\usepackage[linktocpage=true]{hyperref}
\hypersetup{
colorlinks=true,
citecolor=MyDarkBlue,
linkcolor=MyDarkBlue,=
urlcolor=MyDarkBlue,
pdfauthor={},
pdftitle={},
pdfsubject={hep-th}
}

\usepackage[numbers,sort&compress]{natbib}
\usepackage{hypernat}}
\usepackage{graphicx}
\usepackage{cite,color}

\def\e{{\rm e}}\def\ln{{\rm ln}}

\def\a{\alpha}

\def\t{\tilde}
\def\ch{{\rm cosh}}
\def\sh{{\rm sinh}}
\def\cos{{\rm cos}}
\def\sin{{\rm sin}}
\def\d{\partial}

\newcommand{\be}{\begin{equation}}
\newcommand{\bea}{\begin{eqnarray}}

\newcommand{\ee}{\end{equation}}
\newcommand{\eea}{\end{eqnarray}}

\newcommand{\nn}{\nonumber}

\begin{document}

\renewcommand{\thefootnote}{\fnsymbol{footnote}}

\makeatletter
\@addtoreset{equation}{section}
\makeatother
\renewcommand{\theequation}{\thesection.\arabic{equation}}

\rightline{}
\rightline{}


\vspace{10pt}


\begin{center}
{\LARGE \bf{\sc Pulsating strings from two dimensional CFT on $(T^4)^N/S(N)$}}
\end{center} 
 \vspace{1truecm}
\thispagestyle{empty} \centerline{
{\large \bf {\sc Carlos Cardona${}^{a}$}}\footnote{E-mail address: \Comment{\href{mailto:cargicar@ift.unesp.br}}{\tt cargicar@ift.unesp.br}}
 }
\vspace{.5cm}

\centerline{{\it ${}^a$ 
Instituto de F\'{i}sica Te\'{o}rica, UNESP-Universidade Estadual Paulista}} \centerline{{\it  R. Dr. Bento T. Ferraz 271, Bl. II, Sao Paulo 01140-070, SP, Brazil}}

\vspace{1truecm}

\thispagestyle{empty}

\centerline{\sc Abstract}

\vspace{.4truecm}

\begin{center}
\begin{minipage}[c]{380pt}
{We propose a state from the two-dimensional conformal field theory on the orbifold $(T^4)^N/S(N)$ as a dual description for a pulsating string moving in $AdS_3$. We show that, up to first order in the deforming parameter, the energy in both descriptions has the same dependence  on the mode number, but with a non-trivial function of the coupling.
}
\end{minipage}
\end{center}

\vspace{.5cm}

\setcounter{page}{0}
\setcounter{tocdepth}{2}


\renewcommand{\thefootnote}{\arabic{footnote}}
\setcounter{footnote}{0}

\linespread{1.1}
\parskip 4pt

\setcounter{tocdepth}{2}
\tableofcontents

\newpage
\setcounter{equation}{0}

\section{Introduction}

Among the $AdS_{d+1}/CFT_{d}$ dualities \cite{Maldacena:1997re}, one of the most important and studied examples is the correspondence between type IIB string theory on $AdS_3\times S^3\times T^4$ and two-dimensional superconformal field theory.

The duality is motivated from a system of intersecting $Q5$ D5-branes and $Q1$ D1-branes compactified on a four torus $T^4$, whose near horizon geometry is given by $AdS_3\times S^3\times T^4$.
The conjectured dual field theory arises as the infrared
fixed point theory living on the D1-D5 system which can be described by a sigma model given by a deformation of the symmetric product of $N$ copies of a free two dimensional ${\cal N}=(4,4)$ superconformal field theory on the torus $T^4$ or $Sym{(T^4)^N}$ from now on \cite{David:2002wn}, where the $N$ entering the symmetric product is given by $N = Q1 Q5$.

Several aspects of this duality were checked so far. Among them, is worth to mention that after the comparison of the chiral spectrum \cite{Maldacena:1998bw} at both sides, three-point functions of chirals were successfully computed and matched in the supergravity approximation in \cite{Mihailescu:1999cj, Arutyunov:2000by} and beyond supergravity it was done in \cite{Gaberdiel:2007vu,Dabholkar:2007ey,Pakman:2007hn,Cardona:2009hk}. As well as an OPE expansion of four-point functions of chirals were found to agree in both dual theories \cite{Cardona:2010qf}. At first, the equivalence between string theory and field theory correlators was quite remarkable since the computations are performed at different points in the moduli space where solvable descriptions are available \cite{Dijkgraaf:1998gf,Larsen:1999uk}. On the boundary, the solvable point corresponds to a superconformal field theory on $Sym{(T^4)^N}$, while in the string theory it corresponds to a WZW model on $SL(2,\mathbb{R})^2\times SU(2)^2$. A careful analysis of the moduli dependence of the chiral ring eventually
showed though that all three-point functions obey a non-renormalization theorem \cite{deBoer:2008ss,Baggio:2012rr}.

The non-renormalization for half-BPS states, renders the checks done on the chiral spectrum somewhat trivial, and it becomes imperative the study of the duality away from the orbifold point, which would be the equivalent to turning on interactions on a gauge theory. In order to do that, one should include marginal deformations on the CFT in order to move away from the original position in the moduli space. In this respect, few things have been done so far. Classical strings rotating fast along an angular direction of the $S^3$ have been related to chiral-twist fields in the field theory \cite{Gomis:2002qi}, and the energy of quantum excitations about this classical configurations, which can be approximately described in a Penrose limit of the whole background, have been shown to be reproduced from the first correction to the conformal weight of the excited chiral-twist fields in the deformed two-dimensional dual field theory \cite{Gava:2002xb}, and those results were extended for all values of the deforming parameter or coupling in \cite{David:2008yk}. 
However, it is still unclear how the dictionary between states in the two sides of the correspondence works once the deformation is turned on. Very recently, the effect of the marginal deformation in the mixing of certain operators was studied \cite{Burrington:2012yq}. 
Also in an interesting series of papers \cite{Avery:2010er,Carson:2014yxa,Carson:2014xwa}, the effect over the vacuum of the undeformed theory by the action of the marginal deformation was studied, and the authors have found that under the presence of the deformation, the vacuum maps to a squeezed state, very much like the Hawking mechanism of Black Hole particle production \cite{Hawking:1974sw}.

Encouraged by those results, we propose in this short note a type of states in the conformal field theory on the symmetric product to be dual to long pulsating strings in $AdS_3$. We argue that this classical string solutions should correspond to low twist and highly excited states in the $CFT_2$ and we show that the first quantum correction to the conformal weight due to the marginal deformation reproduces the scaling in terms of mode number of the semi-classical energy of quantum excitations around the classical pulsating string, although with a non-trivial function of the coupling.

In the paradigmatic ${\cal N}=4$ SYM duality, large charge operators associated to semi-classical configurations in $AdS_5$ have led to uncovering of integrable structures \cite{Minahan:2002ve, Kazakov:2004qf} since the seminal work of BMN \cite{Berenstein:2002jq}. It is our hope that improving the understanding of the dictionary between semi-classical configurations in $AdS_3\times S^3\times T^4$ and states in the dual conformal field theory away from the orbifold point will lead us to uncover the integrable structures of the boundary theory.
It was only recently shown the worldsheet string integrability in the context of  $AdS_3/CFT_2$ case. Starting with the Bethe-Ansatz proposal in \cite{Babichenko:2009dk}, it was found a weakly coupled spin-chain description of the string \cite{OhlssonSax:2011ms,Sax:2012jv} and an all-loop proposal for the S-matrix for massive modes was given in \cite{Borsato:2013hoa,Borsato:2013qpa} which finally was extended to include the massless modes very recently in \cite{Borsato:2014exa,Borsato:2014hja}. For a nice review on worldsheet integrability of strings on $AdS_3\times S^3\times T^4$ see \cite{Sfondrini:2014via}.

This paper is organized as follows. In section 2 we review classical pulsating string solutions in $AdS_3$ and perform a semi-classical quantization around this classical configurations at two different energy regimes. In section 3 we review some important aspects of the two-dimensional theory which we use in section 4 in order to compute the anomalous dimension of the operators that we proposed as dual to the pulsating string. Finally we devoted section 5 to discussions and conclusions.

\section{Strings in $AdS_3\times S^3\times T^4$}
The near horizon geometry obtained by wrapping $Q_5$ D5-branes on $T^4$ together with $Q_1$ D1-branes parallel to the non-compact direction of the wrapped D5-branes, is given by $AdS_3\times S^3\times T^4$, whose metric can be written as
\bea
ds_{AdS_3}^2&=&R^2(-\ch(\rho)^2dt^2+d\rho^2+\sh(\rho)^2d\phi^2)
\label{AdS3}\,,\\
ds_{S^3}^2&=&R^2(d\theta^2+\cos(\theta)^ 2d\psi^2+\sin(\theta)^2d\varphi^2)\,,\\
ds_{T^4}^2&=&\sqrt{\frac{Q_1}{v Q_5}}\,dx_i^2~~~~~i=1,\cdots,4\,,
\eea
where $R$ is the radius of both $AdS_3$ and $S^3$ and is given by the branes charges as
\be
\frac{R^2}{\alpha'}=g_6\sqrt{Q_1Q_5} 
\ee
where $g_6$ is the effective six-dimensional string coupling and $Q_1Q_5$ should be equal to the multiplicity $N$  of the torus in the symmetric product.

\subsection{Pulsating string in $AdS_3$}
In this section we shall consider a pulsating string moving on $AdS_3$ and static in $S^3\times T^4$ which contracts to a point in $\rho=0$ and then expands up to reach a maximal size $\rho=\rho_{max}$ to contract again and so on. This type of classical motion has been considered in \cite{deVega:1994yz, Gubser:2002tv, Minahan:2002rc} from the perspective of the Nambu-Goto description and from the perspective of the WZW description in \cite{deVega:1998ny, Larsen:1998sq, Maldacena:2000hw} and recently has been revisited for pulsating strings also rotating in $S^3$ \cite{Pradhan:2013sja}, as well as in some deformed backgrounds \cite{Giardino:2011jy}. 
The pulsating string correspond to the following ansatz
\be t=t(\tau),\,\phi=m\sigma,\,\rho=\rho(\tau)\,,\ee
for a string in (\ref{AdS3}) with $R^2/\alpha'=g_6\sqrt{N}\equiv\lambda$. In order to preserve the gravity approximation we should have $R^2>>\a'$, which implies $\lambda>>1$.

In the above ansatz, the Polyakov action reduces to the following,
\be
S=\frac{R^2}{4\pi\alpha'}\int dtd\sigma\left[-\ch^2\rho\,\dot{t}^2+\dot{\rho}^2-m^2\sh^2\rho\right]\,.
\ee The equations of motion are given by,
\bea
2\ch\rho\, \sh\rho\, \dot{\rho}\,\dot{t}+\ch^ 2\rho\,\ddot{t}&=&0\,,\\
\ddot{\rho}+\sh\rho\,\ch\rho\,(m^2+\dot{t}^2)&=&0\label{rhoequat}\,,
\eea
and the Virasoro conditions are
\be\label{vira}
\dot{\rho}^2+m^2\sh^2\rho-\ch^2\rho\,\dot{t}^2=0\,. 
\ee
The conserved energy is obtained as
\be
{\cal E}= \frac{E}{\lambda}=\ch^2\rho\,\dot{t}\,,
\ee
and the conjugate momentum to $\rho$ is $\Pi=2\dot{\rho}$.

Using ${\cal E}$ we can rewrite the Virasoro constraint from (\ref{vira}) as
\be
 \frac{{\cal E}^2}{\ch^ 2\rho}=          \dot{\rho}^2+m^2\sh^2{\rho}\,.
\ee
For small ${\cal E}$, i.e, for energies $E<<\lambda$ the last equation simplifies to 
\be {\cal E}^2=\frac{\Pi^2}{4}+m^2\rho^2\,,\ee
which means that for small energies the spectrum of the string is the square root  of an harmonic oscillator, i.e
\be\label{Esmall}
{\cal E}= \sqrt{m(n+1/2)}\,.
\ee

One can also quantize the string in the large ${\cal E}$ region, or $E>>\lambda$, by following \cite{Minahan:2002rc} we perform a Bohr-Sommerfeld analysis. The quantized states satisfy
\be
n=\frac{\lambda}{2\pi}\oint d\rho\Pi=\frac{\lambda}{\pi}\int_0^{\rho_{max}}d\rho\sqrt{\frac{{\cal E}^2}{\ch^ 2\rho}-m^2\sh^2{\rho}}\,.
\ee

 In \cite{Pradhan:2013sja} this is integral has been performed, and they found for large values of ${\cal E}$ the following
 \be {\cal E} =2n/\lambda+a_1\sqrt{mn/\lambda}+a_2 m-a_3\frac{m^{3/2}\sqrt{\lambda}}{\sqrt{n}}\ee
or,
 \be\label{1loopE}  E =2n+a_1\sqrt{\lambda}\sqrt{mn\pi}+a_2 m-a_3\frac{(m\lambda)^{3/2}}{\sqrt{n}}=2n\left(1+\frac{a_1}{2}\left(\frac{m\lambda}{n}\right)^{1/2}-\frac{a_3}{2}\left(\frac{m\lambda}{n}\right)^{3/2}\right)\,,\ee
where $a_1=8\sqrt{\pi}/\Gamma(1/4)^2\sim 1.07$. This case corresponds to a large string exploring the geometry of $AdS_3$ and we are looking at the space of parameters such that $1<<\lambda<<E$, i.e, $\sqrt{\lambda}/(2n)<<1$ .
 
\section{Review of the field theory}\label{FTR}
The low energy description for a system of $Q_1$ D1-branes and $Q_5$ D5-branes compactified over $\mathbb{R}\times T^4$ is given by a field theory in two dimensions. One can vary the moduli of string theory $(g_s,{\rm Vol}_{T^4},Q_1,Q_5,...)$ to rake a variety of theories at different points. It has been conjectured \cite{Arutyunov:1997gi,deBoer:1998ip,Seiberg:1999xz,Dijkgraaf:1998gf,Larsen:1999uk,Jevicki:1998bm}
that there exist a point in the moduli space where the field theory is particularly simple and is given by a two-dimensional sigma model ${\cal N}=4$ SCFT whose target space is given by a symmetric product of   $N=Q_1Q_5$  copies of $T^4$, 
\be \frac{(T^4)^N}{S(N)}\,,\ee
where $S(N)$ is the permutation group of $N$ elements.

The symmetry algebra of this theory, displayed in the appendix, is generated by the modes of the energy-momentum tensor $L_n, \bar{L}_n$ obeying the Virasoro algebra, plus the modes of the $R-$symmetry group  $SU(2)_R\times SU(2)_L$ denoted by $J_n^a, \bar{J}_n^a$, plus the supersymmetric partners  modes $G^{\pm}_r,\bar{G}^{\pm}_r$.
 
{\it Chiral twist operators}
The action of the symmetric group $S(N)$ is implemented by the following boundary conditions,
\be\label{bouncond}
\d x^i_{I}(z\e^{2\pi i})=\d x^i_{I+1}(z)\,,\quad\text{for}\quad I=1...M,\,\quad i=
1,\cdots,4\,,\ee
where $x^i_I$ are the four coordinates of the $I-$th copy of $T^4$.  
The above twisted boundary conditions can be induced by twisted operators $\sigma_M(w)$ satisfying \cite{Dixon:1986qv, Jevicki:1998bm}
\be
\d x^i_I(z)\sigma_M(w)\sim z^{\frac{1}{M}-1}\e^{-\frac{2\pi i}{M}}\tau_M(w)+\text{ Reg}\,.
\ee
They have conformal weight $\Delta_M=(M-1/M)/24$ and are uncharged under the $R-$symmetry. Chiral operators are built from twisted operators by acting on them with the raising operators $J^+_{n}$ of the $SU(2)-$R symmetry algebra. We denote the chiral operators in twisted sector $M$ by ${\cal O}_M$. Its conformal weight equals its charge under the Casimir of the $SU(2)_L\times SU(2)_R$ $R-$symmetry, given by
\be\label{chiral} \Delta({\cal O}_M)=\frac{M-1}{2}=J\,.\ee

\subsection{Conformal perturbation theory}

The duality between the conformal field theory on the symmetric product and type IIB strings on $AdS_3\times S^3\times T^4$ has been checked so far only at the point in the moduli space given by the two dimensional theory briefly described above. It is our intention in this note to propose a check away from that ``orbifold" point. More precisely, we would like to deform the CFT theory by turning on a marginal deformation given by an operator ${\cal O}_d$ of conformal weight $h=(1,1)$. Those type of operators are singlets under the $SU(2)_L\times SU(2)_R$ $R-$symmetry, they preserve the ${\cal N}=(4,4)$ supersymmetry and are constructed by applying the super-current modes $G^{aA}_{-1/2},\,\bar{G}^{\dot{a}B}_{-1/2}$ to the twist-two primary operator $\sigma_{2}$ with conformal dimension $(1/2,1/2)$. They have the schematic form
\be\label{defope}
{\cal O}_d\sim \epsilon_{AB}G^{+ A}_{-1/2}\t{G}^{+B}_{-1/2}\sigma_{2}^{++}\,.
\ee
Given the deforming operator, we add the following deformation to the action
\be S_{int}=\t{\lambda}\int d^2z{\cal O}_d(z,\bar{z})+ {\rm a.c}\ee
By performing Kadanoff's conformal perturbation theory analysis\cite{Kadanoff:1979}, it is found that the first correction to the two-point function is given by
\be \langle\phi^i(z_1)\phi^j(z_2) \rangle_{\t{\lambda}}=\delta^{ij}
\left(1-2\t{\lambda}\frac{\d h^i}{\d\t{\lambda}}\ln(z_{12})\right)\langle\phi^i(z_i)\phi^j(z_j) \rangle_{0}\,,\ee
where $h^i,\,h^j$ are the conformal weights of $\phi^i,\,\phi^j$ respectively. Therefore, by computing the two point function we will be able to read the correction to the conformal weights due to the deformation. By using path integral together with $SL(2,\mathbb{C})$ invariance, it has been computed in \cite{Eberle:2001jq,Burrington:2012yq} that

\be\label{w1loop}\frac{\d h_i}{\d\t{\lambda}}=-\pi C_{idi}\,,\ee
where
\be\label{3p1}
C^{idi}=z_0\lim_{z_3\to\infty}z_3\langle\phi^i(0){\cal O}_d(z_0)\phi^i(z_3) \rangle_0\,.
\ee

\section{Large occupancy number states}
In this section we will propose a field theory dual to the classical pulsating string in $AdS_3$.
It is already well known \cite{Gubser:2002tv} that the pulsating string should be dual to a highly excitated state, i.e, from the perspective of the two-dimensional CFT in the symmetric product we expect a state with a large occupation number $m>>1$. 
Our proposal for states dual to pulsating strings is the following. Since the pulsating string is static in the $S^3$, the state is not charged under the  $SU(2)_L\times SU(2)_R$ $R-$symmetry currents of the CFT model, i.e, it should have $J=0$, which in terms of the chiral fields satisfying (\ref{chiral}) means $M=1$. On the other hand, the classical string is highly excited in $AdS_3$, whose isometries correspond to the Virasoro generators in the boundary, i.e, the state should have large $(L_0,\, \bar{L}_0)$, but at the same time, since the string is not rotating, the state should satisfy $L_0-\bar{L}_0=0$ and therefore we should associate the energy of the string with $E=2 L_0$. Specifically, we propose the dual operator to be given by the following descendant state,
\be\label{dualstate}
|n\rangle\equiv \frac{L_{-n}}{2^{n/2}}|J=0\rangle\,,
\ee
created from the asymptotic state 
\be
|J\rangle=\lim_{z\to 0}{\cal O}_J(z)|0\rangle\,. 
\ee
The energy of the state (\ref{dualstate}) in the undeformed theory $E=2L_0=2n$, coincides with the classical energy of the pulsating string (\ref{1loopE}). 

From equation (\ref{w1loop}) and (\ref{3p1}), we compute,
\bea\frac{\d h_n}{\d\t{\lambda}}=-\frac{\pi}{2^{n}} z_0\lim_{z_3\to\infty}z_3\langle (L_n{\cal O}_{})(z_3){\cal O}_d(z_0)(L_{-n}{\cal O}_{1})(0)\rangle_0\,.
\eea
Since we are interested only in the scaling of the one-loop correction as a function of $n$, not caring  much about the exact coefficient, we restrict the computation to the bosonic contribution of the deformation operator, i.e, we drop the fermionic modes multiplying $\sigma_2$ in (\ref{defope}), i.e, 
\be\label{w1loop2}
\frac{\d h_n}{\d\t{\lambda}}=-\frac{\pi\,z_0}{2^{n}}\langle L_n\,\sigma_2(z_0)\,L_{-n}\rangle\,. 
\ee
In order to compute the three point function in (\ref{w1loop2}), we will use the technique developed in \cite{Lunin:2000yv,Lunin:2001pw}, which consists in mapping the computation from the multivalued $z-$plane to a covering plane where the fields are single valued and hence there are no twist insertions.   
Specifically, we have to consider the following picture: The operator creating the asymptotic state $|J=0\rangle$ in the $z-$plane is inserted at the origin,  with $J$ given by (\ref{chiral}). The deformation twist-two operator is located at position $z_0$ creating a branch point of order two in the $z-$plane. 

In order to implement the boundary conditions (\ref{bouncond}) on the energy-momentum tensor
\be 
T_{I}(z\e^ {2\pi i})=T_{I+1}(z)\,,\quad\text{for}\quad I=1...M\,,
\ee
we should use the following expansion under the presence of a twist-$M$ operator \cite{Jevicki:1998bm,Dixon:1986qv},
\be
T_I(z)=-i\sum_n L_n z^{-\frac{n}{M}-2}\e^{-2\pi I\frac{n}{M}}\,,
\ee
where
\be\label{viraz}
L^{(z)}_n=i\oint_{0\leq{\rm Arg}(z)< 4\pi}\, dz\, z^{\frac{n}{M}+1}\e^{2\pi\frac{ n}{M}I}T_I(z)\,.
\ee 
Since we want to consider the correction to the conformal weight of (\ref{dualstate}) due to the presence of the twist-two operator, we need to take particularly $M=2$. Then we have,
\be
T_I(z)=-i\sum_n L_n z^{-\frac{n}{2}-2}\,,
\ee
\be\label{viraz}
L^{(z)}_n=i\oint_{0\leq{\rm Arg}(z)< 4\pi}\, dz\, z^{\frac{n}{2}+1}T_I(z)\,.
\ee 
 In order to have the correlators be single valued, we perform a covering map
\be\label{map1} z=t(t-1)\,, \ee 
which has the required monodromies, with $z_0=1/4$.

Since we will end up with a free field theory on the $t-$plane, it is natural to define the Virasoro modes there as \cite{Avery:2010er}
\be
L^{(t)}_n=\oint\, dt\, t^{n+1}T(t)\,,
\ee
given that, by transforming (\ref{viraz}) from the $z-$plane to the $t-$plane, the modes are related in the following way
 \bea\label{modetrans}
L^{(z)}_n=i\sum_{p\geq0}\left( \begin{array}{c}
n/2+1\\
p\\
\end{array} \right)\,(-1)^pL^{(t)}_{n+1-p}\,.
\eea
Then the first correction to the conformal weight is given by (\ref{w1loop2})
\bea\label{2pLargm}
&&-\frac{\pi\,z_0}{2^{n}}\langle L^{(z)}_n\,\sigma_2(z_0)\,L^{(z)}_{-n}\rangle~~~(z\to t)\nn\\
&=&-\frac{\pi\,z_0}{2^{n}}\left\langle\left(\sum_{p\geq0}\left( \begin{array}{c}
n/2+1\\
p\\
\end{array} \right)(-1)^pL^{(t)}_{n+1-p}\right)\left(\sum_{q\geq0}\left( \begin{array}{c}
n/2+1\\
q\\
\end{array} \right)(-1)^q L^{(t)}_{q-n-1}\right) \right\rangle\,,\nn\\
&=&-{\cal N}\frac{2\pi\,z_0}{2^{n}}\sum_{q\geq0}^{n+1}\left( \begin{array}{c}
n/2+1\\
q\\
\end{array} \right)^2(n+1-q)\,.
\eea
In the above expression we have defined ${\cal N}$ as the expectation value $\langle0|L_0|0\rangle={\cal N}$, which should be equal to the vacuum energy in the $t-$space times the normalization of the two-point functions $\langle0|0\rangle$. Since the theory in the $t-$space is given by a single copy of the conformal field theory on $T^4$, consistent with four scalars plus four fermions, the central charge in it is given by $c_t=(1+1/2)4=6$, then the vacuum energy is given by $c_t/24=1/4$. Additionally, it has been shown in \cite{Lunin:2001pw, Lunin:2000yv} that true CFT operators should be normalized by a factor 
\be
\lambda_M=\left[\frac{N!}{M(N-M)!}\right]^{1/2}\,,
\ee
which in the case we are considering, $M=1\, (J=0)$,  should be $\sqrt{N}$, i.e, ${\cal N}=\sqrt{N}/4$.


The sum in the last line of (\ref{2pLargm}) is given by
\bea
\sum_{q\geq0}^{n+1}\left( \begin{array}{c}
n/2+1\\
q\\
\end{array} \right)^2(n+1-q)=(2+3n)\frac{(n+2)!}{[2(n/2+1)!]^2}+\left(\begin{array}{c}
n/2+1\\
n+2\\
\end{array} \right)^2 F(n),\nn\\
\eea
where $F(n)$ is a finite function as $n$ goes to infinity\footnote{$_3F_2[1,n/2+1,n/2+1|n+3,n+3|1]$}. For large values of $n$, the binomial coefficient multiplying $F(n)$ goes like $(1/n!)^2$, and the first term on the right hand side is approximately given by
\be
(2+3n)\frac{(n+2)!}{[2(n/2+1)!]^2}\sim3\frac{2^{n+2}}{\sqrt{8\pi}}\sqrt{n}\,.                                                                                                                                                                                                                                                                                                                                                                                                                                                                                                                                                                                                                                   \ee
It has been argued in \cite{Gava:2002xb,Gomis:2002qi} that one should relate the deformation parameter $\t{\lambda}$ with the effective string coupling  $g_6$, meaning that in the deformed field theory the 't Hooft parameter is  $\lambda=\t{\lambda}\sqrt{N}$, which should be small for the computation we are performing. 
Putting all together we end up with 
\be 
\frac{\d h_n}{\d \t{\lambda}}=3\frac{{\cal N}{8\pi}}{4\sqrt{8\pi}}\sqrt{n}=\left(\frac{3\sqrt{8\pi} }{16}\right)\sqrt{N}\sqrt{n\pi}\,.
\ee
Therefore the conformal weight of (\ref{dualstate}) up to first order in perturbation theory is given by
\be h_n=h^{(0)}_n+\t{a}_1{\lambda}\sqrt{\pi n}\,, \ee
with $\t{a}_1=\left(\frac{3\sqrt{8\pi} }{16}\right)$, hence the energy of the state up to first order in small $\lambda$ due to the deformation is equal to
\be\label{1loopEweak}
E_{\t{\lambda}}=2n\left(1+ \frac{\t{a}_1}{2}\lambda\sqrt{\frac{\pi}{n}}\right)\,.
\ee
Numerically, $\t{a}_1=0.94$. 

Several comments are in order. 
Since we are interested in the large $n$ limit and the classical  energy\footnote{Or undeformed at the field theory side} for both descriptions are proportional to $n$, we should have the condition that the energy is large at both sides, and then we expected to match  (\ref{1loopEweak}) with the high energy behaviour of the string (\ref{1loopE}). 
As we just see, the leading correction to the conformal weight of the proposed stated goes like $\lambda/\sqrt{n}$, which can be interpreted as an effective expansion parameter for small 'tHoof coupling $\lambda$ and large mode number $n$.  On the other hand, from equation (\ref{1loopE}), we see that the first correction to the energy of the string when ${\cal E}$ is large, is controlled (for $m=1$) by the parameter $\sqrt{\lambda}/\sqrt{n}$ which can be kept small for large $n$ no matter the value of $\lambda$.
Aside of this, the equation (\ref{Esmall}) is valid whenever ${\cal E}$ is small, or  $\lambda\sqrt{n}<<1$  (for $m=1$), which is hard to maintain  for arbitrarily large values of $n$ and therefore we do not expect to describe this behaviour from the field theory for large $n$.

From the discussion above it looks like the genuine parameter to do perturbation at large coupling for this type of configurations is given by $\sqrt{\lambda/n}$ while for small $\lambda$ seems to be $\lambda/\sqrt{n}$ instead. Unlikely the  BMN case \cite{Berenstein:2002jq}, where the parameter $\lambda/J^2$ controls the perturbation at both strong and weak coupling,  the case considered here seems to indicate that the parameter controlling the perturbative expansion for the state (\ref{dualstate})
is given by $\frac{h(\lambda)}{\sqrt{n}}$, where the function $h(\lambda)$ intertwine from $h(\lambda)\sim \lambda$ for $\lambda$ very small and $h(\lambda)\sim \sqrt{\lambda}$ for large $\lambda$.
A similar behaviour has been found in $AdS_4/ABJM$ for the rotating string in ${\mathbb CP}^3$ \cite{Gaiotto:2008cg,Nishioka:2008gz} as well as for the open string \cite{Cardona:2014ora}. At weak 't Hooft coupling the effective parameter is given by $\lambda^2/J^2$ whereas at strong 't Hooft coupling is $\lambda/J^2$.

Although we do not hope to match the coefficients $a_1$ in (\ref{1loopE}) and $\t{a}_1$ in (\ref{1loopEweak}) at large and weak coupling respectively, we found remarkable that they are very close to each other and approximately equal to one. We think it could be indicating that the fermionic contributions to the deforming parameter does not play an important role in the specific computation we have considered.
\section{Discussion and conclusions}
In this paper we have proposed an operator in the two-dimensional conformal field theory on the symmetric product of $N-$copies of $T^4$, as a dual weak coupling description of a pulsating string in $AdS_3$. In the high energy limit, we have found that the first leading order correction to the energy of both configurations (state and string) behave as $h(\lambda)/\sqrt{n}$, where the function $h(\lambda)$  is proportional to $\sqrt{\lambda}$ for large coupling  whereas is given by $\lambda$  for small 't Hooft coupling.

It should be nice to obtain the next-to-leading order correction to the energy of the string (\ref{1loopE}) from the conformal field theory, which implies the insertion of more than one twist-two  deformation parameter. The difficulty for higher order corrections is that, since the additional insertions, one should consider correlation functions of order higher than three, which unlikely three-point functions like (\ref{3p1}), depends on moduli variables on the sphere with punctures which have to be integrated over.

It also would be nice to look for the meaning at the field theory side of the winding number $m$ in the string, which is currently unclear for us.

We hope our results, and in general the study of operators corresponding to classical strings in $AdS_3\times S^3\times T^4$, help to improve the understanding of integrability in the conformal field theory on the symmetric product $Sym(T^4)^N$ after turning on the deforming parameter.

\section*{Acknowledgements}

This work is supported in part by CNPq grant 160022/2012-6. I would like to thank to Horatiu Nastase and Victor Rivelles for reading and correct the manuscript. 
\appendix

\section{${\cal N}=(4,4)$ superconformal algebra.}

The ${\mathcal N}= (4,4)$ SCFT on $(T^4)^N/S(N)$ 
is described by the free Lagrangian
\be
\label{freen4}
S = \frac{1}{2} \int d^2 z\; \left[\d
x^i_A \bar\d x_{i,A} + 
\psi_A^i(z) \bar\d \psi^i_A(z) + 
\widetilde\psi^i_A(\bar z) \d \widetilde \psi^i_A(\bar z) 
 \right]
\ee
Here $i$ runs over the $T^4$ coordinates
1,2,3,4 and $A=1,2,\ldots,N=Q_1Q_5$ labels various copies
of the four-torus. The symmetric group $S(N)$
acts by permuting the copy indices. It introduces the twisted
sectors which we have discussed in section \ref{FTR}.

The algebra is generated by the stress energy tensor $T(z)$, four supersymmetry currents $G^a(z), G^{b\dagger}(z)$, and a local $SU(2)$ $R$ symmetry
current $J^i(z)$. The operator product expansions(OPE) of the algebra 
with central charge $c$ are given by 
\bea
\label{chp2:scft-algebra}
T(z)T(w) &=& \frac{\d T(w)}{z-w} + \frac{2 T(w)}{(z-w)^2} +
\frac{c}{2 (z-w)^4},  \\  \nonumber
G^a(z)G^{b\dagger }(w) &=& 
\frac{2 T(w)\delta_{ab}}{z-w} + \frac{2 \bar{\sigma}^i_{ab} \d J^i}
{z-w} + \frac{ 4 \bar{\sigma}^i_{ab} J^i}{(z-w)^2} + 
\frac{2c\delta_{ab}}{3(z-w)^3}, \\
\nonumber
J^i(z) J^j(w) &=& \frac{i\epsilon^{ijk} J^k}{z-w} + \frac{c}{12 (z-w)^2}
, \\  \nonumber
T(z)G^a(w) &=& \frac{\d G^a (w)}{z-w} + \frac{3 G^a (z)}{2 (z-w)^2},
\\   \nonumber
T(z) G^{a\dagger }(w) &=& \frac{\d G^{a\dagger} (w)}{z-w} + 
\frac{3 G^{a\dagger} (z)}{2 (z-w)^2}, \\   \nonumber
T(z) J^i(w) &=& \frac{\d J^i (w)}{z-w} + \frac{J^i}{(z-w)^2}, \\
\nonumber
J^i(z) G^a (w) &=& \frac{G^b(z) (\sigma^i)^{ba}}{2 (z-w)}, \\
\nonumber
J^i(z) G^{a\dagger}(w) &=& -\frac{(\sigma^i)^{ab} G^{b\dagger
}(w)}{2(z-w)} 
\eea
The $\sigma$'s stand for Pauli matrices and the
$\bar{\sigma}$'s stand for the complex conjugates of Pauli matrices.
The above OPE generates an affine ${\mathcal N}=4$ superconformal algebra, whose global part forms the
supergroup $SU(1,1|2)$. 
Let $L_{\pm,0} ,J^{(1),(2),(3)}_R$  be
the global charges of the currents
$T(z)$ and $J^{(i)}_R(z)$ and  $G^a_{1/2,-1/2} $  the
global charges of the supersymmetry currents $G^a(z)$ 
in the Neveu-Schwarz sector. From the OPE's 
we obtain the following commutation relations for the global charges.
\bea
[L_0, L_{\pm}] = \mp L_{\pm} \;\;&\;&\;\; [L_{1} , L_{-1}] = 2L_{0} 
\\ \nonumber
\{ G^a_{1/2} , G^{b\dagger}_{-1/2} \} &=& 2\delta^{ab}L_0 + 2
\sigma^i_{ab} J^{(i)}_{R} \\  \nonumber
\{ G^a_{-1/2} , G^{b\dagger}_{1/2} \} &=& 2\delta^{ab}L_0 - 2
\sigma^i_{ab} J^{(i)}_{R} \\  \nonumber
[J^{(i)}_R, J^{(j)}_R] &=& i\epsilon^{ijk}J^{(k)}_R \\ \nonumber
[L_0, G^a_{\pm 1/2}] = \mp\frac{1}{2} G^a_{\pm 1/2} \;\;&\;&\;\;[L_0, G^{a\dagger}_{\pm 1/2}] = \mp\frac{1}{2} G^{a\dagger}_{\pm 1/2} 
\\ \nonumber
[L_+ , G^a_{1/2}] = 0 \;\;&\;&\;\; [L_- , G^a_{-1/2}] = 0 \\ \nonumber
[L_- , G^a_{1/2}] = -G^a_{-1/2} \;\;&\;&\;\; [L_+ , G^a_{-1/2}] = G^a_{1/2} \\ \nonumber
[L_+ , G^{a\dagger}_{1/2}] = 0 \;\;&\;&\;\; [L_- , G^{a\dagger}_{1/2}] = 0 \\ \nonumber
[L_- , G^{a\dagger}_{1/2}] = -G^a_{-1/2} \;\;&\;&\;\; [L_+ , G^{a\dagger}_{-1/2}] = G^a_{1/2} \\ \nonumber
[J^{(i)}_R, G^a_{\pm 1/2} ] = \frac{1}{2} G^{b}_{\pm 1/2} (\sigma^i)^{ba}
\;\;&\;&\;\; [J^{(i)}_R, G^{a\dagger}_{\pm 1/2} ] = -\frac{1}{2}
 (\sigma^i)^{ba}G^{b\dagger}_{\pm 1/2} \\  \nonumber
\eea

The representations of the supergroup $SU(1,1|2)$
are classified according to the conformal weight 
and $SU(2)_R$ quantum number. \\The highest weight states
$ |h\rangle = 
|h,{\bf j}_R=j,j_R^3 =j \rangle $ satisfy the following
properties
\bea
L_1 |h\rangle = 0 &\;\;\; 
L_0 |h\rangle = h|h\rangle \\ \nonumber
J^{(+)}_{R}|h\rangle =0  &\;\;\; 
J_R^{(3)}|h\rangle = j_R|h\rangle\\  \nonumber
G_{1/2}^a|h\rangle =0 &\;\;\; G_{1/2}^{a\dagger}
|h\rangle =0
\eea
where $J^+_R = J^{(1)}_R + i J^{(2)}_R$.
Highest weight states which satisfy 
$
G^{2\dagger }_{-1/2}|h\rangle =0 ,\;\;\;
G^1_{-1/2}|h\rangle =0
$
are chiral primaries. They satisfy $h=j$. We will denote 
these states as $|h\rangle _{S}$. Short multiplets are
generated from the chiral primaries through the action of the raising
operators $J^{-}, G^{1\dagger }_{-1/2}$ and $G^2_{-1/2}$. 
\bibliography{EAB}{}
\bibliographystyle{utphys}

\end{document}